\documentclass[conference,a4paper]{IEEEtran}
\usepackage{cite}
\ifCLASSINFOpdf
\usepackage[pdftex]{graphicx}
\graphicspath{./figures}
\DeclareGraphicsExtensions{.pdf,.jpeg,.png}
\else
\fi
\usepackage[cmex10]{amsmath}
\usepackage{array}
\usepackage{subcaption}
\usepackage{url}
\usepackage[utf8]{inputenc}  
\usepackage{citesort}
\usepackage{bigints}
\usepackage{float}
\usepackage[T1]{fontenc} 
\usepackage{mathtools}
\usepackage{amsfonts}
\usepackage{amssymb}
\usepackage{algorithm}
\usepackage[noend]{algpseudocode}
\usepackage{multirow}
\algrenewcommand{\algorithmiccomment}[1]{\hfill$\triangleright$ #1}
\newcolumntype{M}[1]{>{\centering\arraybackslash}m{#1}}
\newcolumntype{N}{@{}m{0pt}@{}}
\usepackage[font={small}]{caption}
\DeclareUnicodeCharacter{00A0}{~}

\def \({\left(}
\def \){\right)}
\def \[{\left[}
\def \]{\right]}
\newcommand{\tbf}[1]{{\textbf{#1}}}

\newcommand{\defeq}{\vcentcolon=}

\newcommand{\bA}{{\textbf {A}}}

\newcommand{\bx}{{\textbf {x}}}

\newcommand{\by}{{\textbf {y}}}

\newcommand{\bs}{{\textbf {s}}}

\newcommand{\be}{\begin{equation}}
\newcommand{\ee}{\end{equation}}
\newcommand{\bea}{\begin{eqnarray}}
\newcommand{\eea}{\end{eqnarray}}

\begin{document}
\title{Generalized Approximate Message-Passing Decoder for Universal Sparse Superposition Codes}
\author{\IEEEauthorblockN{Erdem Bıyık$^*$, Jean Barbier$^{\dagger}$ and Mohamad Dia$^{\dagger}$\\
$*$ Bilkent University, Ankara, Turkey \\
$\dagger$ Communication Theory Laboratory, EPFL, Lausanne, Switzerland\\
erdem.biyik@ug.bilkent.edu.tr, \{jean.barbier, mohamad.dia\}@epfl.ch}}
\maketitle
\IEEEpeerreviewmaketitle
\begin{abstract}
Sparse superposition (SS) codes were originally proposed as a capacity-achieving communication scheme over the additive white Gaussian noise channel (AWGNC) \cite{barron2010sparse}. Very recently, it was discovered that these codes are universal, in the sense that they achieve capacity over any memoryless channel under generalized approximate message-passing (GAMP) decoding \cite{barbierDiaMacris_thresholdSat}, although this decoder has never been stated for SS codes. In this contribution we introduce the GAMP decoder for SS codes, we confirm empirically the universality of this communication scheme through its study on various channels and we provide the main analysis tools: state evolution and potential. We also compare the performance of GAMP with the Bayes-optimal MMSE decoder. We empirically illustrate that despite the presence of a phase transition preventing GAMP to reach the optimal performance, spatial coupling allows to boost the performance that eventually tends to capacity in a proper limit. We also prove that, in contrast with the AWGNC case, SS codes for binary input channels have a vanishing error floor in the limit of large codewords. Moreover, the performance of Hadamard-based encoders is assessed for practical implementations.
\end{abstract}
\section{Introduction}
Sparse superposition codes were introduced by Barron and Joseph for communication over the AWGNC \cite{barron2010sparse}. These codes were proven to achieve the Shannon capacity using power allocation and various efficient decoders \cite{JosephB14,barron2012high}. 
A decoder based on approximate message-passing (AMP), originally developed for compressed sensing \cite{Donoho10112009,KrzakalaMezard12}, was introduced in 
\cite{barbier2014replica}. The authors in \cite{rush2015capacity} proved, using the state evolution (SE) analysis \cite{bayati2011dynamics,Javanmard21102013}, that AMP allows to achieve 
capacity using power allocation. 
At the same time, spatially coupled SS codes were introduced in \cite{barbierSchulkeKrzakala,BarbierK15} and empirically shown 
to approach capacity under AMP without power allocation and to perform much better than power allocated ones. Recently, AMP for spatially coupled SS codes was shown to saturate the so-called {\it potential threshold}, related to the Bayes-optimal MMSE performance, which tends to capacity in a proper limit \cite{barbierDiaMacris_isit2016}.

This set of works combined with the excellent performance of SS codes over the AWGNC motivated their study for any memoryless channel under GAMP decoding \cite{barbierDiaMacris_thresholdSat}. GAMP was introduced as a generalization of AMP for generalized estimation \cite{rangan2011generalized}. In \cite{barbierDiaMacris_thresholdSat} the authors showed that, under the assumption that SE \cite{Javanmard21102013} tracks GAMP for SS codes, spatially coupled SS codes achieve the capacity of any memoryless channel under GAMP decoding. However, GAMP has never been explicitly stated or tested as a decoder for SS codes other than for the AWGNC, in which case GAMP and AMP are identical. In this work we fill this gap by studying the GAMP decoder for SS codes over various memoryless channels. We focus on the AWGNC (for completeness with previous studies \cite{barbierSchulkeKrzakala,BarbierK15}), binary erasure channel (BEC), Z channel (ZC) and binary symmetric channel (BSC). However, the present decoder and analysis remain valid for any memoryless channel.

Our experiments confirm that SE recursion of \cite{barbierDiaMacris_thresholdSat} accurately tracks GAMP. Using the potential of the code we also compare the performance of GAMP to the optimal MMSE decoder. In addition, our empirical study confirms the asymptotic results of \cite{barbierDiaMacris_thresholdSat}: the performance of SS codes under GAMP decoding can be significantly increased towards capacity using spatial coupling, as already observed for the AWGNC \cite{BarbierK15}. Moreover, we prove that for binary input channels, SS codes have a vanishing error floor in the limit of large codewords even with finite sparsity. This means that when decoding is possible, optimal decoding is asymptotically perfect as well as GAMP decoding until some threshold, a very promising feature which is {\it not} present for the real-valued input AWGNC. Keeping in mind practicality, we focus our empirical study on Hadamard-based coding operators that allow to drastically reduce the encoding and decoding complexity, while maintaining good performance for moderate block-lengths \cite{barbierSchulkeKrzakala}.
\section{Sparse superposition codes: Setting}
% In the sequel, a Gaussian random variable $X$ with mean $m$ and variance $\sigma^2$ is denoted $X \!\sim\! \mathcal{N}(m,\sigma^2)$ and the corresponding density $\mathcal{N}(x\vert m, \sigma^2)$.
%
In SS codes, the \emph{message} $\bx \!=\! [\bx_1, \dots, \bx_L]$ is a vector made of $L$ $B$-dimensional \emph{sections}. Each section $\bx_l$, $l\!\in\!\{1,\ldots,L\}$, satisfies a hard constraint: it has a single non-zero component
equals to $1$ whose position encodes the symbol to transmit. $B$ is the \emph{section size} (or alphabet size) and we set $N\!\defeq\!LB$. For the theoretical analysis we consider random codes generated by a \emph{coding matrix} $\bA\!\in\! \mathbb{R}^{M \times N}$ drawn from the ensemble of Gaussian matrices with i.i.d entries $\sim\!{\cal N}(0,\sigma_\bA^2)$. For the practical implementation, fast Hadamard-based operators are used instead as they exhibit very good performances. Despite the lack of rigor in the analysis for such operators, they remain good predictive tools \cite{barbierSchulkeKrzakala}. The \emph{codeword} is $\bA\bx\!\in\! \mathbb{R}^{M}$. We enforce the power constraint $||\bA\bx||_2^2/M\!=\!1$ by tuning $\sigma_\bA^2$. The cardinality of the code is $B^L$. Hence, the (design) rate is $R\!=\!L\log_2(B)/M$ and the code is thus specified by $(M, R, B)$.

The aim is to communicate through a known memoryless channel $W$. This requires to map the continuous-valued codeword onto the input alphabet of $W$. The concatenation of this mapping operation and the channel itself can be interpreted as an \emph{effective memoryless channel} $P_{\text{out}}(\by|\bA\bx)\!=\!\prod_{\mu=1}^M P_{\text{out}}(y_\mu|[\bA\bx]_\mu)$. For the channels we focus on, $P_{\text{out}}(y_\mu|[\bA\bx]_\mu)$ is expressed as follows:
\begin{itemize}
	\item AWGNC: $\mathcal{N}(y_\mu|[\bA\bx]_\mu, 1/{\rm snr})$,
	\item BEC: $(1\!-\!\epsilon)\delta(y_\mu\!-\!{\rm sign}([\bA\bx]_\mu))\! +\! \epsilon\delta(y_\mu)$,	
	\item BSC: $(1\!-\!\epsilon)\delta(y_\mu\!-\!{\rm sign}([\bA\bx]_\mu)) \!+\! \epsilon\delta(y_\mu\!+\!{\rm sign}([\bA\bx]_\mu))$,
	\item ZC: $\delta({\rm sign}([\bA\bx]_\mu)\!+\!1)(\epsilon\delta(y_\mu\!-\!1)\!+\!(1\!-\!\epsilon)\delta(y_\mu\!+\!1))\!+\!\delta({\rm sign}([\bA\bx]_\mu)\!-\!1)\delta(y_\mu\!-\!1)$,
\end{itemize}
where $\textrm{snr}$ is the signal-to-noise of the AWGNC, $\epsilon$ the erasure or flip probability of the BEC, ZC and BSC. The $\rm{sign}$ maps the Gaussian distributed codeword components onto the input alphabets of the binary input channels. 

Note that for the asymmetric ZC, the symmetric map ${\rm sign}([\bA\bx]_\mu)$ leads to a sub-optimal uniform input distribution. The {\it symmetric capacity} of the ZC differs from Shannon's capacity but the difference is small, and similarly for the algorithmic threshold, see \cite{barbierDiaMacris_thresholdSat}. We thus consider this symmetric setting for the sake of simplicity. The other channels are symmetric, this map thus leads to the optimal input distribution.

%
%\begin{figure}[h]
%	\centering
%	\includegraphics[width=0.45\textwidth]{./figures/blockDiag_encoder_decoder}
%	\vspace*{-2pt}
%	\caption{Scheme for the use of GAMP decoder algorithm. GAMP regards everything after the encoder as the effective channel. HD in the schematic stands for hard decision. In hard decision step, the results of the MMSE decoder GAMP, $\hat{x}$, are mapped to SS code $\bar{x}$ such that the largest value in each section is set to 1, and the rest to 0.}
%	\label{fig:blockDiag}
%	\vspace{-10px}
%\end{figure}
%
% We now shortly present the \emph{spatially coupled} ensemble of SS codes: Spatially coupled matrices $\in\!\mathbb{R}^{M\times N}$ are made of $(\Gamma\!+\!1)\!\times \!\Gamma$ blocks. This ensemble of matrices is parametrized by $(M,R,B,\Gamma_R,\Gamma_C,w_{left},w_{right},g_w)$, where $w_{left}$ and $w_{right}$ are the \emph{coupling windows} and $g_w$ is the \emph{design function}. The detailed construction is explained in \cite{barbierDiaMacris_isit2016}.
%
\section{The GAMP decoder}\label{sec:gamp}
\begin{algorithm}[t!]
		\caption{{GAMP}{\,($\textbf{y},\bA,B,t_{\rm max},u$})}\label{alg:gamp}
		\begin{algorithmic}[1]
			\State $\widehat{\bx}_{(0)}= \boldsymbol{0}_{N,1}$, ${\boldsymbol{\tau}^x}_{(0)}= (1/B)\boldsymbol{1}_{N,1}$		
			\State ${\bs}_{(-1)} = \boldsymbol{0}_{M,1}$, $t= 0$, $e_{(0)}= \infty$	\Comment{Initializations}
			\While{$t\le t_{\rm max}$ \textbf{and} $ e_{(t)}\ge u$}
			\State $\boldsymbol{\tau}^p_{(t)}\hspace{10pt}= \hspace{7pt}\bA^{\circ2}{\boldsymbol{\tau}^x_{(t)}}$ 
			\State $\textbf{p}_{(t)}~~~~\!\!\hspace{0.5pt}= \hspace{7pt}\bA\widehat{\bx}_{(t)} - {\boldsymbol{\tau}^p_{(t)}}\circ{\bs_{(t-1)}}$\Comment{Output linear step}
			\State $\boldsymbol{\tau}^s_{(t)}~~~= \hspace{7pt}-g_{\text{out}}'(\textbf{p}_{(t)}, \textbf{y}, \boldsymbol{\tau}^p_{(t)})$
			\State $\bs_{(t)}~~~~~~\!\!\!\!\!\hspace{0.5pt}= \hspace{7pt}g_{\text{out}}(\textbf{p}_{(t)}, \textbf{y}, \boldsymbol{\tau}^p_{(t)})$\Comment{Output non-linear step}		
			\State $\boldsymbol{\tau}^r_{(t)}~~~= \hspace{7pt}({((\boldsymbol{\tau}^s_{(t)})^\intercal\bA^{\circ2})^\intercal})^{\circ-1}$
			\State $\textbf{r}_{(t)}~~~~~\!\!\!=\hspace{7pt}\widehat{\bx}_{(t)} + {\boldsymbol{\tau}^r_{(t)}}\circ({\bs}_{(t)}^\intercal\bA)^\intercal$\Comment{Input linear step}
			\State $\boldsymbol{\tau}^x_{(t+1)}\hspace{0.5pt}= \hspace{7pt}\boldsymbol{\tau}^r_{(t)}\circ g_{\text{in}}'(\textbf{r}_{(t)}, \boldsymbol{\tau}^r_{(t)})$
			\State $\widehat{\bx}_{(t+1)}~~\!\!\!= \hspace{7pt}g_{\text{in}}(\textbf{r}_{(t)}, \boldsymbol{\tau}^r_{(t)})$\Comment{Input non-linear step}	
			\State $e_{(t)} \hspace{13pt}= \hspace{7pt}||\widehat{\bx}_{(t+1)} - \widehat{\bx}_{(t)}||_2^2/L$	
			\State $t\hspace{23.5pt}= \hspace{7pt}t+1$			
			\EndWhile\label{gampwhile}
			\State \textbf{return} $\widehat{\bx}_{(t)}$\Comment{The prediction scores for each bit}			
		\end{algorithmic}		
\end{algorithm}
We consider a Bayesian setting and associate to the message the posterior $P(\bx|\by,\bA) \!=\! P_{\text{out}}(\by|\bA\bx) P_0(\bx)/P(\by|\bA)$. The hard constraints for the sections are enforced by the prior $P_0(\bx)\!=\! \prod_{l=1}^L p_0(\bx_l)$ with $p_0(\bx_l)\!=\!B^{-1}\sum_{i\in l} \delta_{x_i,1}\prod_{j\in l, j\neq i} \delta_{x_j,0}$, where $\{i\!\in\!l\}$ are the $B$ scalar components indices of the section $l$. The GAMP decoder aims at performing MMSE estimation by approximating the posterior mean of each section. 

In the GAMP decoder Algorithm~\ref{alg:gamp}, $\circ$ denotes element-wise operations. GAMP was originally derived for {\it scalar} estimation. In this generalization to the vectorial setting of SS codes, whose derivation is similar to the one of AMP for SS codes found in \cite{BarbierK15}, only the input non-linear steps differ from canonical GAMP \cite{rangan2011generalized}: here the so-called \emph{denoiser} $g_{\text{in}}$ acts {\it sectionwise} instead of componentwise. In full generality, it is defined as \cite{rangan2011generalized} $g_{\text{in}}(\textbf{r},{\boldsymbol{\tau}})\!\defeq\!\mathbb{E}[\textbf{X} | \textbf{R}\!=\!\textbf{r}]$ for the random variable $\textbf{R}\!=\!\textbf{X}\!+\! \widehat{\textbf{Z}}$ with $\textbf{X}\!\sim\!P_0$ and $\widehat{\textbf{Z}}\!\sim\!{\cal N}(0,\text{diag}({\boldsymbol{\tau}}))$. Moreover, the estimate of the posterior variance, which quantifies how ``confident'' GAMP is in its current estimate, equals $\boldsymbol{\tau}~\! \circ~\! g_{\text{in}}'(\textbf{r}, {\boldsymbol{\tau}})\!=\!\mathbb{E}[\textbf{X}^{\circ2} | \textbf{R}\!=\!\textbf{r}]\!-\!g_{\text{in}}(\textbf{r}, {\boldsymbol{\tau}})^{\circ2}$ ($g_{\text{in}}'$ is the componentwise partial derivative w.r.t its first argument, and similarly for $g_{\text{out}}'$). Plugging $P_0$ yields the componentwise expression of the denoiser and the variance term:
\begin{align*}
	\begin{cases}
	[g_{\text{in}}(\textbf{r}, {\boldsymbol{\tau}})]_i& = \frac{\exp((2r_i-1)/(2 \tau_i))}{\sum_{j\in l_i}\exp((2r_j-1)/(2 \tau_j))}, \nonumber\\
	[\boldsymbol{\tau}\! \circ \!g_{\text{in}}'(\textbf{r}, {\boldsymbol{\tau}})]_i &=[g_{\text{in}}(\textbf{r}, {\boldsymbol{\tau}})]_i(1\!-\![g_{\text{in}}(\textbf{r}, {\boldsymbol{\tau}})]_i), \nonumber
	\end{cases}
\end{align*}
$l_i$ being the section to which belong the $i^{\text{th}}$ scalar component.

In contrast with $g_{\text{in}}$ that only depends on $P_0$, $g_{\text{out}}$ depends on the communication channel and acts componentwise. Its general form and specific expressions for the studied channels are given in Table~\ref{table:gampForSS} along with the necessary derivatives.
\begin{table*}[t]
	\caption{The expressions for $g_{\rm out}$, $-g_{\rm out}'$ and $\mathcal{F}$.}\label{table:gampForSS}
	\centering
	{
	\vspace*{-5pt}
	\begin{tabular}{|M{0.95cm}|M{4.2cm}|M{7.95cm}|M{2.60cm}|N}
	\hline
	&
	$[g_{\rm out}(\textbf{{p}},\textbf{y},\boldsymbol{\tau})]_i$ & $[\!-\!g_{\rm out}'(\textbf{{p}}, \textbf{y}, \boldsymbol{\tau})]_i$
	& $\mathcal{F}(p|E)$
	&\\[5pt]
	\hline
	\textbf{General} & 
	$(\mathbb{E}[Z_i | p_i,y_i, \tau_i] \!-\! {p}_i)\!/\!{\tau}_i$
	$Y_i\sim P_{\rm out}(\cdot|z_i), Z_i\sim \mathcal{N}(p_i,\tau_i)$
	& $({{\tau}_i\!-\! \textrm{Var}[Z_i|p_i,y_i,\tau_i]})\!/\!{\tau}_i^2$

	$Y_i\sim P_{\rm out}(\cdot|z_i), Z_i\sim \mathcal{N}(p_i,\tau_i)$
	& See \eqref{Fisher}
	& \\[8pt]
	\hline
	\textbf{AWGNC} & 
	$\frac{{y}_i\!-\!p_i}{{\tau}_i\!+\!1\!/\!\textrm{snr}}$
	& $\frac1{{\tau}_i\!+\!1\!/\!\textrm{snr}}$
	&
	$\frac1{1\!/\!\textrm{snr}\!+\!E}$
	&\\[8pt]
	\hline
	\textbf{BEC} & 
	$\frac{(p_i\!-\!k_i)h^+_i\!+\!(p_i\!+\!k_i)h^-_i\!+\!2\epsilon\delta({y}_i)p_i}{{\cal Z}_{\textrm{BEC}}\tau_i}\!-\!\frac{p_i}{\tau_i}$
	& $\frac1{{\tau}_i}\!-\!\frac{(p_i^2\!+\!{\tau}_i\!-\!k'_i)h^+_i\!+\!(p_i^2\!+\!{\tau}_i\!+\!k'_i)h^-_i\!+\!2\epsilon\delta({y}_i)({\tau}_i\!+\!p_i^2)}{{\cal Z}_{\textrm{BEC}}{{\tau}_i}^2}\!+\!\big(\![g_{\rm out}\!(\!\textbf{{p}},\!\textbf{y},\!\boldsymbol{\tau}\!)]_i\!+\!\frac{p_i}{{\tau}_i}\!\big)\!^2$
	&
	$\frac{Q'^2(1\!-\!\epsilon)}{Q(1\!-\!Q)}$
	&\\[8pt]
	\hline
	\textbf{ZC} & 
	$\frac{(p_i\!-\!k_i)v^+_i\!+\!(p_i\!+\!k_i)\delta({y}_i\!-\!1)}{{\cal Z}_{\textrm{ZC}}\tau_i}\!-\!\frac{p_i}{\tau_i}$
	& $\frac1{{\tau}_i}\!-\!\frac{(p_i^2\!+\!{\tau}_i\!-\!k'_i)v^+_i\!+\!(p_i^2\!+\!{\tau}_i\!+\!k'_i)\delta({y}_i\!-\!1)}{{\cal Z}_{\textrm{ZC}}{{\tau}_i}^2}\!+\!\big(\![g_{\rm out}\!(\!\textbf{{p}},\!\textbf{y},\!\boldsymbol{\tau}\!)]_i\!+\!\frac{p_i}{{\tau}_i}\!\big)\!^2$
	& 
	$\frac{Q'^2(1\!-\!\epsilon)^2}{Q\!+\!\epsilon(1\!-\!Q)}\!+\!\frac{Q'^2(1\!-\!\epsilon)}{1\!-\!Q}$
	&\\[8pt]
	\hline
	\textbf{BSC} & 
	$\frac{(p_i\!-\!k_i)v^+_i\!+\!(p_i\!+\!k_i)v^-_i}{{\cal Z}_{\textrm{BSC}}\tau_i}\!-\!\frac{p_i}{\tau_i}$
	& $\frac1{{\tau}_i}\!-\!\frac{(p_i^2\!+\!{\tau}_i\!-\!k'_i)v^+_i\!+\!(p_i^2\!+\!{\tau}_i\!+\!k'_i)v^-_i}{{\cal Z}_{\textrm{BSC}}{{\tau}_i}^2}\!+\!\big(\![g_{\rm out}\!(\!\textbf{{p}},\!\textbf{y},\!\boldsymbol{\tau}\!)]_i\!+\!\frac{p_i}{{\tau}_i}\!\big)\!^2$
	& 
	$\frac{Q'^2(1\!-\!2\epsilon)^2}{(Q\!+\!\epsilon\!-\!2\epsilon Q)(1\!-\!Q\!-\!\epsilon\!+\!2\epsilon Q)}$
	&\\[8pt]
	\hline
	\multicolumn{4}{|c|}{$h^+_i\!=\!(1\!\!-\!\!\epsilon)\delta({y}_i\!\!+\!\!1)$,\quad $h^-_i\!=\!(1\!\!-\!\!\epsilon)\delta({y}_i\!\!-\!\!1)$,\quad $v^+_i\!=\!(1\!\!-\!\!\epsilon)\delta({y}_i\!\!+\!\!1)\!\!+\!\!\epsilon\delta({y}_i\!\!-\!\!1)$,\quad $v^-_i\!=\!(1\!\!-\!\!\epsilon)\delta({y}_i\!\!-\!\!1)\!\!+\!\!\epsilon\delta({y}_i\!\!+\!\!1)$,}
	&\\[8pt]
	\multicolumn{4}{|c|}{ ${k}_i\!=\!\textrm{exp}\big(\frac{-p_i^2}{2\tau_i}\big)\sqrt{2{{\tau}_i}/\pi}\!\!+\!\!\textrm{erf}\big(\frac{p_i}{\sqrt{2{{\tau}_i}}}\big)p_i$,\quad $k'_i\!=\!{k}_i p_i\!\!+\!\!\textrm{erf}\big(\frac{p_i}{\sqrt{2{{\tau}_i}}}\big){\tau}_i$,\quad $Q \!=\! \frac12 \textrm{erfc}(\frac{\!-\!p}{\sqrt{2E}})$, \quad $Q'\!=\!\textrm{exp}\big(\frac{-p^2}{2E}\big)\big/{\sqrt{2\pi E}}$}
	&\\[8pt]
	\multicolumn{4}{|c|}{${\cal Z}_{\textrm{BEC}}\!=\! {\textrm{erfc}\big(\frac{p_i}{\sqrt{2{\tau}_i}}\big)h^+_i\!\!+\!\!\big(1\!\!+\!\!\textrm{erf}\big(\frac{p_i}{\sqrt{2{\tau}_i}}\big)\big)h^-_i\!\!+\!\!2\epsilon\delta({y}_i)}$, ${\cal Z}_{\textrm{ZC}}\!=\!\textrm{erfc}\big(\frac{p_i}{\sqrt{2{\tau}_i}}\big)v^+_i\!\!+\!\!\big(1\!\!+\!\!\textrm{erf}\big(\frac{p_i}{\sqrt{2{\tau}_i}}\big)\big)\delta({y}_i\!\!-\!\!1)$, ${\cal Z}_{\textrm{BSC}}\!=\!\textrm{erfc}\big(\frac{p_i}{\sqrt{2{\tau}_i}}\big)v^+_i\!\!+\!\!\big(1\!\!+\!\!\textrm{erf}\big(\frac{p_i}{\sqrt{2{\tau}_i}}\big)\big)v^-_i$}
	&\\[8pt]
	% \multicolumn{4}{|c|}{${\cal Z}_{\textrm{BSC}}\!=\!\textrm{erfc}\big(\frac{p_i}{\sqrt{2{\tau}_i}}\big)v^+_i\!+\!\big(1\!+\!\textrm{erf}\big(\frac{p_i}{\sqrt{2{\tau}_i}}\big)\big)v^-_i$}
	% &\\[8pt]	
	\hline
	\end{tabular}}
	\vspace{-10px}
\end{table*}

The complexity of GAMP is dominated by the $\mathcal{O}(MN)\!=\!\mathcal{O}(L^2B\ln(B))$ matrix-vector multiplications. In terms of memory, it is necessary to store $\bA$ which can be problematic for large codes. Fast Hadamard-based operators constructed as in \cite{barbierSchulkeKrzakala}, with random sub-sampled modes of the full Hadamard operator, allow to achieve a lower $\mathcal{O}(L\ln(B) \ln(BL))$ decoding complexity and strongly reduce the memory need \cite{BarbierK15,condo_practical}.
\section{State evolution and the potential}\label{sec:stateandpot}
\begin{figure}[t]
	\centering
	\vspace*{-5pt}
	\includegraphics[width=0.47\textwidth]{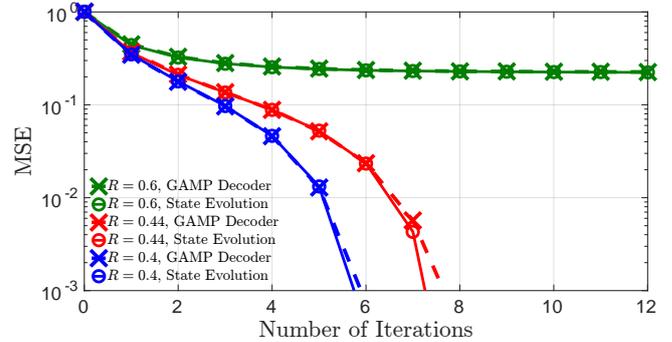}
	% \vspace*{-4pt}
	\caption{SE tracking the GAMP decoder (averaged over $100$ random instances) over the BEC with erasure probability $\epsilon\!=\!0.1$ and for $L\!=\! 2^{11}$, $B \!=\! 4$ and Gaussian coding operators. Monte carlo integration with $2\!\times\! 10^{4}$ samples is used for the computation of SE. The algorithmic threshold $R_{\rm GAMP}\!\approx \!0.55$ and the green curves are for a rate above it: decoding fails. In contrast, the blue and red curves are below $R_{\rm GAMP}$: decoding succeeds. After the last points of these curves, both SE and the GAMP curves fall to $0$ MSE.}
	\label{fig:se_track}
	\vspace{-15pt}
\end{figure}
We now present the analysis tools of the $L\!\to\!\infty$ performance of SS codes under GAMP and MMSE decoding when Gaussian matrices are used: state evolution and potential.
\subsection{State evolution}
The asymptotic performance of GAMP with Gaussian i.i.d coding matrices is tracked by SE,
a scalar recursion \cite{bayati2011dynamics,Javanmard21102013,rangan2011generalized,barbierDiaMacris_thresholdSat} analogous to density evolution for low-density parity-check codes. Note that although SE is not rigorous for vectorial setting, the rigorous analysis of \cite{rush2015capacity} and the present empirical results strongly suggest that {\it it is exact}, which we conjecture.

The aim is to compute the asymptotic MSE of the GAMP estimate $E_{(t)} \!\defeq\! \lim_{L \to\infty }\|\widehat{\bx}_{(t)} \!-\! \bx\|_2^2/L$. It turns out that this is equivalent to recursively compute the MMSE $T(E)\! \defeq \!\mathbb{E}_{\textbf{S},\textbf{Z}}[\|\textbf{S}\!-\! \mathbb{E}[\textbf{X}|\textbf{S}\!+\!(\Sigma(E)/b)\textbf{Z}]\|_2^2]$ of a single section ($\textbf{S}\!\sim\! p_0$) sent through an equivalent AWGNC ($\textbf{Z}\!\sim\! \mathcal{N}(0,\tbf{I}_B)$) of noise variance $(\Sigma(E)/b)^2$, $b^2\!\!\defeq\!\log_2(B)$. This formulation is valid for any memoryless channel \cite{barbierDiaMacris_thresholdSat}, $P_{\rm out}$ being reflected in
\begin{align}
	\begin{cases}
	\Sigma(E) &\defeq \sqrt{R}[\int dp\, \mathcal{N}(p|0,1\!-\!E)\mathcal{F}(p|E)]^{-1/2},\\
	\mathcal{F}(p|E) &\defeq \int dy f(y|p,E) (\partial_x \ln f(y|x,E))^2_{x=p},	\label{Fisher}\\
	f(y|p,E) &\defeq \int dz P_{\text{out} }(y|z) \mathcal{N}(z|p,E).
	\end{cases}
\end{align}
$\mathcal{F}$ is the Fisher information of $p$ associated with $f$, see Table~\ref{table:gampForSS}. The $p$ integral in $\Sigma$ can be numerically computed for the BEC, ZC and BSC. Define  
\begin{align*}
	\begin{cases}
	g_{\text{in}}^{(1)}(\Sigma,\textbf{z}) &\defeq \big[1\!+\! e^{-\frac{b^2}{\Sigma^2}}\sum_{j=2}^B e^{\frac{b}{\Sigma}(z_j-z_1)}\big]^{-1},\\
	g_{\text{in}}^{(2)}(\Sigma,\textbf{z}) &\defeq \big[1\!+\!e^{\frac{b^2}{\Sigma^2}+(z_1-z_2)\frac{b}{\Sigma}}\!+\!\sum_{k=3}^B{e^{(z_k-z_2)\frac{b}{\Sigma}}}\big]^{-1}.
	\end{cases}
\end{align*}
The MMSE of the equivalent AWGNC is obtained after simple algebra \cite{BarbierK15} and reads $$T(E) \!=\! \mathbb{E}_\textbf{Z}[(g_{\text{in}}^{(1)}(\Sigma(E),\textbf{Z})\!-\!1)^2\!+\!(B\!-\!1)g_{\text{in}}^{(2)}(\Sigma(E),\textbf{Z})^2].$$ Here $g_{\text{in}}^{(1)}$ is interpreted as the posterior mean approximated by GAMP of the non-zero component in the transmitted section while $g_{\text{in}}^{(2)}$ corresponds to the remaining components. The SE recursion tracking the MSE of GAMP is then
\begin{align}
E_{(t+1)} = T(E_{(t)}), \quad t\geq 0,	\label{SE}
\end{align}
initialized with $E_{(0)}\!=\!1$. Hence, the asymptotic MSE reached by GAMP upon convergence is $E_{(\infty)}$. Moreover, define the asymptotic $L\!\to\!\infty$ {\it error floor} of SS codes $E_*$ as the fixed point of SE \eqref{SE} initialized from $E_{(0)}\!=\!0$.
Fig.~\ref{fig:se_track} shows that SE properly tracks GAMP on the BEC. Note that the section error rate (SER) of GAMP, the fraction of wrongly decoded sections after hard thresholding of $\widehat{\bx}_{(t)}$, can also be asymptotically tracked thanks to SE through a simple one-to-one mapping between $E_{(t)}$ and the asymptotic SER at $t$ \cite{barbier2014replica,BarbierK15}.

Under GAMP decoding SS codes exhibit, as $L\!\to\!\infty$, a sharp phase transition at an \emph{algorithmic threshold} $R_{\rm GAMP}$ below Shannon's capacity. $R_{\rm GAMP}$ is defined as the highest rate such that for $R\!\le\!R_{\rm GAMP}$, \eqref{SE} has a unique fixed point $E_{(\infty)}\!=\!E_*$ (see \cite{barbierDiaMacris_thresholdSat} for formal definitions). In this regime GAMP decodes well, see red and blue curves of Fig.~\ref{fig:se_track}.
% If \eqref{SE} has multiple fixed points, $E_*$ is the one associated with the global min. of the potential, see next section, while we always denote $E_{(\infty)}$ the one asymptotically reached by GAMP; i.e the fixed point of SE with initialization $E_{(0)}\!=\!1$. 
If $R\!>\!R_{\rm GAMP}$ GAMP decoding fails, see green curve. As we will see in the next sections, spatial coupling may allow to boost the performance of the scheme by increasing the GAMP algorithmic threshold.
\subsection{Potential formulation}\label{subsec:potentials}
The SE \eqref{SE} is associated with a {\it potential} $F_{\rm u}(E)$, whose {\it stationary points correspond to the fixed points of SE}: $\partial_EF_{\rm u}(E)|_{E_0}\!=\!0\!\Leftrightarrow\!T(E_0)\!=\!E_0$. For SS codes it is \cite{barbierDiaMacris_thresholdSat}:
\begin{align*}
	\begin{cases}
	F_{\rm u}(E) &\defeq U_{\rm u}(E) \!-\! S_{\rm u}(\Sigma(E))\\
	U_{\rm u}(E) &\defeq -\frac{E}{2\ln(2)\Sigma(E)^2}\! -\! \frac{1}{R} \mathbb{E}_Z[\int dy\, \phi \log_2(\phi)], \\
	%S_{\rm u}( \Sigma( {E})) \defeq \mathbb{E}_{\mathbf{z}}[\log_B(\int d{\mathbf{s}} \,P_0(\mathbf{s}) \theta(\mathbf{s},\mathbf{z},\Sigma(E)))],\\
	S_{\rm u}( \Sigma( {E})) &\defeq \mathbb{E}_{\textbf{Z}}\bigr[\log_B\big(1+\sum_{i=2}^B e_{i}(\textbf{Z},\Sigma(E)/b)\big)\bigl],
	\end{cases}
\end{align*}
where $\phi\!=\!\phi(y|Z,E)\! \defeq \!\int ds P_{\text{out}}(y|s)\mathcal{N}(s|Z \sqrt{1 \!-\! E}, E)$, $Z\!\sim\!\mathcal{N}(0,1)$ and $e_{i}(\textbf{Z},x)\! \defeq\! \exp\big((Z_i\!-\!Z_1)/x\!-\!1/x^2 \big)$. 

It has been recently shown for random linear estimation, including compressed sensing and SS codes with AWGN \cite{BDMK_alerton2016,ReevesPfister_isit16}, that $\min_{E\in[0,1]}F_{\rm u}(E)$ equals the asymptotic mutual information (up to a trivial additive term) and that {\it $\widetilde E\!\defeq\!{\rm argmin}_{E\in[0,1]}F_{\rm u}(E)$ equals the asymptotic MMSE}. A proof for all memoryless channels remains to be done, but we conjecture that it remains true under mild conditions on $P_{\rm out}$.
 
\begin{figure}[t]
	\centering
	\vspace*{-1.5pt}
	\includegraphics[trim={0 0 0 0}, clip,width=0.45\textwidth]{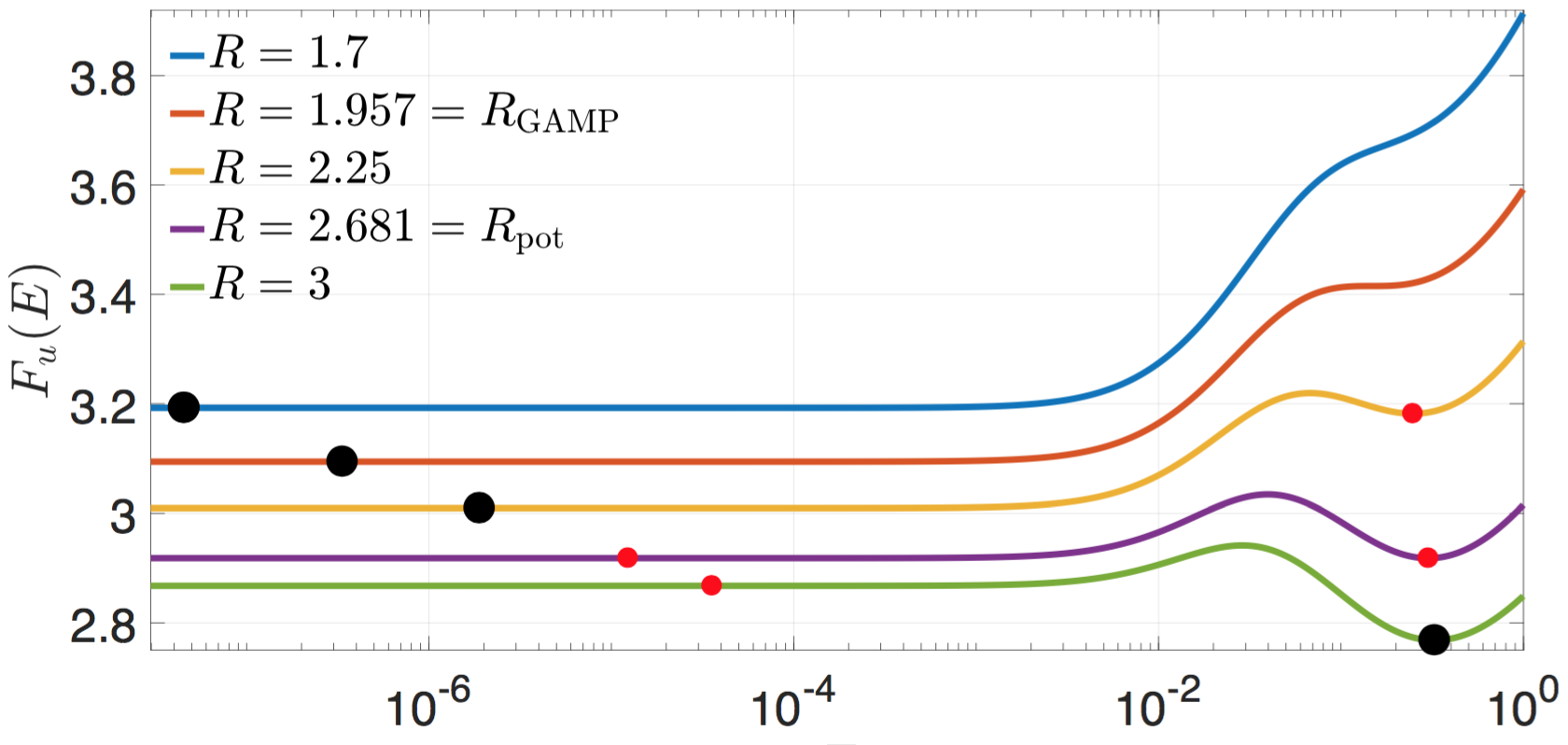}
	\includegraphics[width=0.45\textwidth]{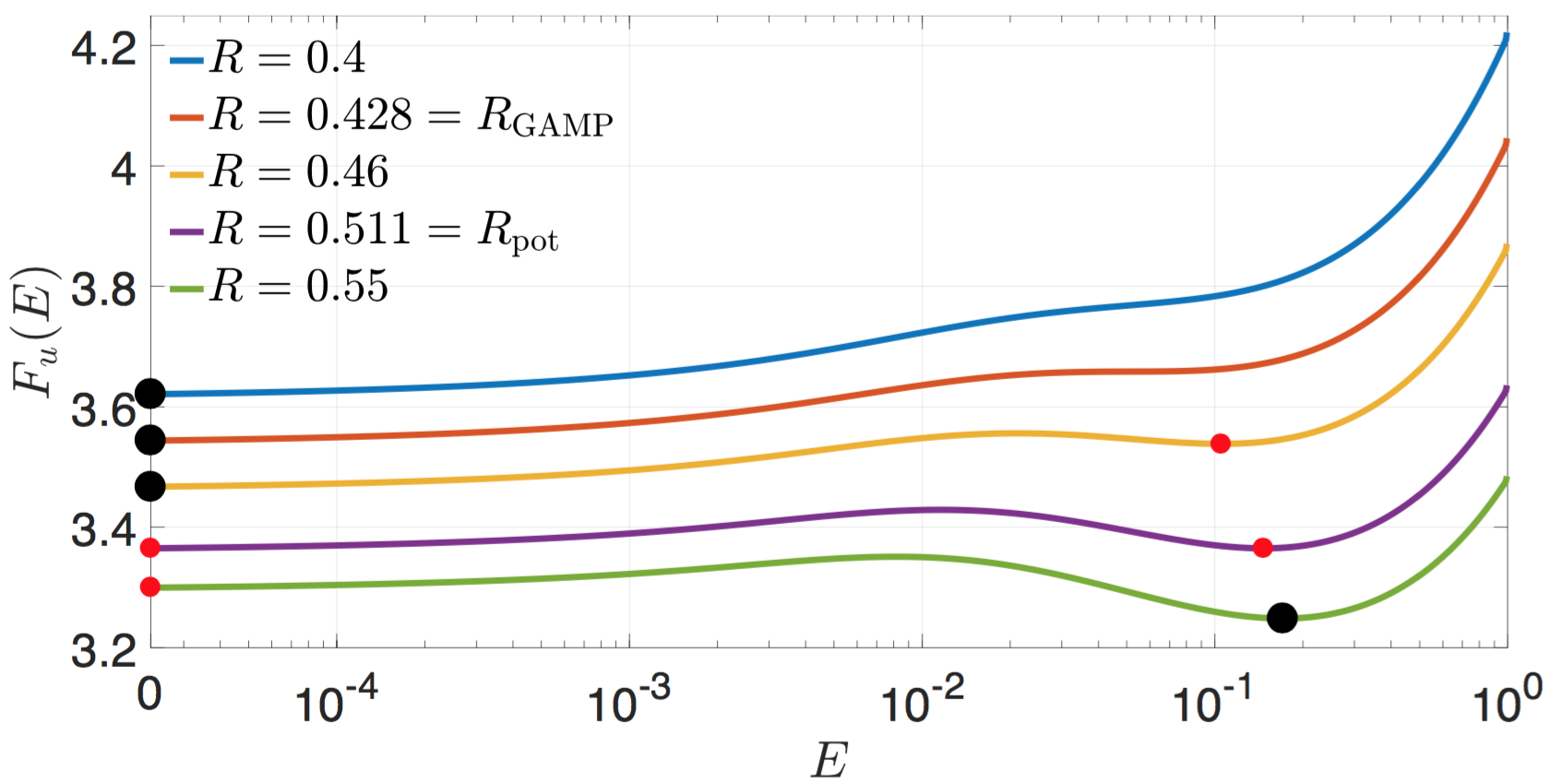}	
	\hspace*{-4pt}
	\caption{Potential for the AWGNC with $\rm{snr}\!=\!100$ (top) and the BEC with $\epsilon\!=\!0.1$ (bottom), in both cases with $B \!=\!2$. The MMSE is the ${\rm argmin}\,F_{\rm u}(E)$ (black dot). When the min. is unique (i.e $R\!<\!R_{\rm GAMP}$, blue curve) or if the global min. is the rightmost one ($R\!>\!R_{\rm pot}$, green curve), GAMP is asymptotically optimal, despite that if $R\!>\!R_{\rm pot}$ it leads to poor results. The red dot is the local min., preventing GAMP to decode if $R\!\in]R_{\rm GAMP}, R_{\rm pot}[$ (yellow curve).}
	\label{fig:potential}
	\vspace{-16pt}
\end{figure}

Using these properties of the potential and its link with SE, it is possible to assess the performances of the GAMP and MMSE decoders by looking at its minima. GAMP decoding is possible (and asymptotically optimal as it reaches the MMSE $\widetilde E$, black dot in Fig.~\ref{fig:potential}) for rates lower or equal to $R_{\rm GAMP}$, whose equivalent definition is the smallest solution of $\partial F_{\rm u}/ \partial E\!=\! \partial^2 F_{\rm u}/\partial E^2 \!=\! 0$; in other words it is the smallest rate at which a horizontal inflection point appears in the potential, see blue and red curves in Fig.~\ref{fig:potential}. For $R\!\in]R_{\rm GAMP},R_{\rm pot}[$, referred to as the {\it hard phase}, the potential possesses another local min. (red dot) and the corresponding ``bad'' fixed point of SE prevents GAMP to reach $\widetilde E$; decoding fails (yellow curves). Finally, the rate at which the local and global min. switch roles is the \emph{potential threshold} $R_{\rm{pot}}$ (purple curves). Optimal decoding is possible as long as $R\!<\!R_{\rm pot}$ as the MMSE switches at $R_{\rm pot}$ from a ``low'' to a ``high'' value. At higher rates GAMP is again optimal but leads to poor results as decoding is impossible. Note that if $R\!<\!R_{\rm pot}$, then $E_*\!=\!\widetilde E$.
\\ 

In the hard phase, where two minima coexist, spatial coupling enables decoding \cite{barbierSchulkeKrzakala} by ``effectively suppressing'' the spurious local min. of the potential. It implies that the algorithmic threshold of spatially coupled SS codes $R^{\rm c}_{\rm{GAMP}}$, the highest attainable rate using coupled codes under GAMP decoding \cite{barbierDiaMacris_thresholdSat}, saturates the potential threshold $R_{\rm pot}$ in the limit of infinite coupled chains. This phenomenon is referred to as {\it threshold saturation} and is understood as the generic mechanism behind the excellent performances of coupled codes \cite{6589171,barbierDiaMacris_thresholdSat}. Moreover, a very interesting feature of SS codes is that $R_{\rm pot}$ itself approaches the capacity as $B\!\to\!\infty$ \cite{barbierDiaMacris_thresholdSat}. These phenomena imply together that in these limits (infinite chain length and $
B$), spatially coupled SS codes under GAMP decoding are {\it universal} in the sense that they achieve the Shannon capacity of all memoryless channels.
\subsection{Vanishing error floor for binary input memoryless channels} \label{sec:noError}
Another promising feature of SS codes is related to their error floor. In the real-valued input AWGNC case, an error floor always exists but it can be made arbitrary small by increasing $B$ \cite{BarbierK15,barbierDiaMacris_thresholdSat}: $\lim_{B\to\infty}E_*\!=\lim_{B\to\infty}\text{SER}_*\!=\!0$, $\text{SER}_*$ the error floor in the SER sense. In contrast, in the BEC, ZC and BSC cases (more generally for binary input memoryless channels), we now prove that {\it as $L\!\to\!\infty$ the error floor vanishes for any $\epsilon$ and $B$}. This implies that when $E_*\!=\!\widetilde E$ optimal decoding is asymptotically perfect, and thus GAMP decoding as well for $R\!\le\!R_{\rm GAMP}$. This is actually verified in practice for GAMP where perfect decoding is statistically possible even for moderate block-lengths, see blue and red curves of Fig.~\ref{fig:se_track}.

The proof of $E_*\!=\!0$, i.e the existence of the {\it trivial fixed point} $T(0)\!=\!0$ of \eqref{SE}, does not guarantee that this is the global minimum of the potential in the hard phase; i.e it is {\it a priori} possible that $E_*\!\neq\!\widetilde E$. Nevertheless, our careful numerical work indicates that {\it there exist at most two fixed points of SE at the same time or equivalently two minima in the potential}, namely $E_*\!=\!\widetilde E\!\neq\!E_{(\infty)}$ if $R\!\in]R_{\rm GAMP}, R_{\rm pot}[$ or $E_*\!\neq\!\widetilde E\!=\!E_{(\infty)}$ if $R\!>\!R_{\rm pot}$ (at least for the studied cases), see Fig.~\ref{fig:potential}. This also agrees with the $B\!\to\!\infty$ analysis of the potential \cite{BarbierK15,barbierDiaMacris_thresholdSat}.

Let us now prove that $E_*\!=\!0$ for the BEC, the proof for other binary input channels being similar. It starts by noticing, from the definition of $T(E)$ as the MMSE of an AWGNC with noise parameter $\Sigma(E)$, that a sufficient condition for $T(0)\!=\!0$ is $\lim_{E\to0}\Sigma(E)\!=\!0$; indeed no noise implies vanishing MMSE. From \eqref{Fisher} this condition is equivalent to $\lim_{E\to0}I_{\mathbb{R}}(E) \!=\!\infty$ that we now prove, where $I_{\mathcal{A}}(E)\!\defeq\!\int_{\mathcal{A}} dp \mathcal{N}(p|0,1\!-\!E)\mathcal{F}(p|E)$. Consider instead $I_{\mathcal{E}}(E)$ where $\mathcal{E}\!\defeq\![E\!-\!\sqrt{E},E\!+\!\sqrt{E}]$. Using Table~\ref{table:gampForSS} for the expression of $\mathcal{F}(p|E)$ for the BEC, this restricted integral is
\begin{align*}
I_{\mathcal{E}}(E)\!=\!\frac{(1\!-\!\epsilon)(2\pi)^{-3/2}}{E\sqrt{1\!-\!E}}\int_{\mathcal{E}} dp \frac{e^{-\frac{p^2}{2(1-E)}-\frac{p^2}{E}}}{Q(p,E)(1\!-\!Q(p,E))}. \label{eq3}
\end{align*}
Here $Q(p,E)\!\in\![C_E,1\!-\!C_E]$, with $\lim_{E\to0}C_E\! >\!0$ for $p\!\in\!\mathcal{E}$, $E\!\le\!1$. This implies that $K(E)\!\defeq\!\max_{p\in \mathcal{E}}Q(p,E)(1\!-\!Q(p,E))\!=\!\mathcal{O}(1)$. Since the interval $\mathcal{E}$ is of size $2\sqrt{E}$, then
\begin{equation}\label{diver}
I_\mathcal{E}(E)\! \ge\! \frac{(1\!-\!\epsilon)(2\pi)^{-3/2}2\sqrt{E}}{E\sqrt{1\!-\!E}K(E)} e^{-\frac{(E+\sqrt{E})^2}{2(1-E)}-\frac{(E+\sqrt{E})^2}{E}}. 
\end{equation}
From this we can assert that $\lim_{E\to0}I_\mathcal{E}(E)\!=\!\infty$. Moreover $I_{\mathcal{E}}(E)\!<\!I_\mathbb{R}(E)$ as $\mathcal{F}(p|E) \!\ge\!0$ (recall it is a Fisher information) and thus $\lim_{E\to0}I_\mathbb{R}(E)\!=\!\infty$ which ends the proof. 

For the BSC and ZC the proof is similar, the main ingredient being the squared Gaussian $Q'^2$ at the numerator of $\mathcal{F}(p|E)$, see Table~\ref{table:gampForSS}, which leads to similar expressions as \eqref{diver} and thus the $1/\sqrt{E}$ divergence when $E\!\to\! 0$. We believe that the same mechanism holds for any binary input memoryless channel, implying a vanishing error floor as well as asymptotic perfect decoding of GAMP below the algorithmic threshold.
\begin{figure}[t!]
	\centering
	\includegraphics[trim={0 0 0 0}, clip, width=0.45\textwidth]{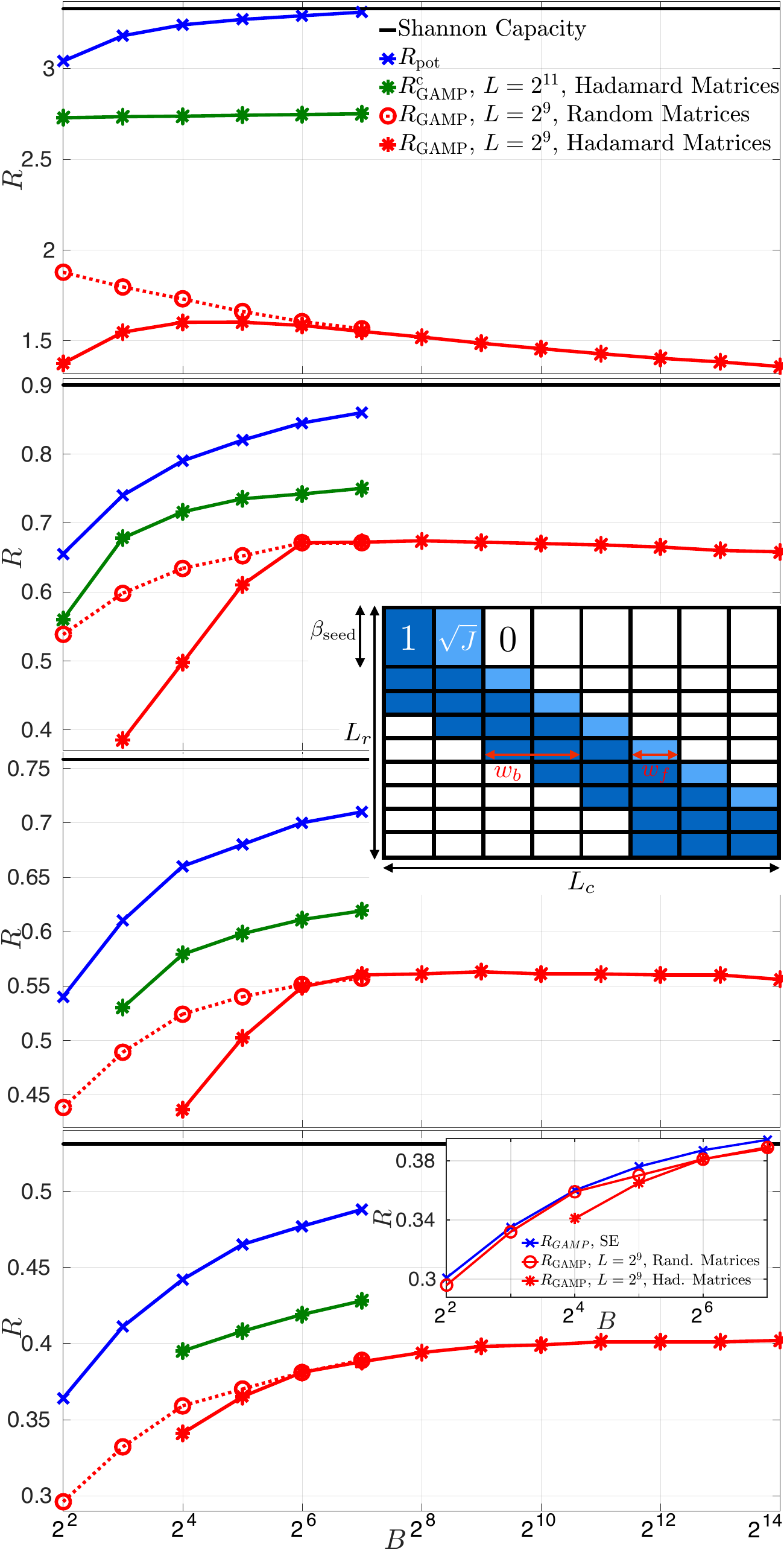}	
	\caption{Phase diagrams for (from top) the AWGNC with ${\rm snr}\!=\!100$, the BEC, ZC and BSC all with $\epsilon\!=\!0.1$. The $L\!\to\!\infty$ transition $R_{\rm pot}$ is obtained from the potential by equating its two minima. $R_{\rm{GAMP}}$, formally defined for $L\!\to\!\infty$, is instead obtained for finite $L\!=\!2^9$ by running GAMP over $100$ instances for each $(R,B)$ and by {\it defining} the transition as the highest rate for which at least $50$ instances were decoded (up to a small SER due to finite size effects). The inner figure illustrates finite size effects by comparing $R_{\rm GAMP}$ computed in this way (for the BSC) and the ``true'' $L\!\to\!\infty$ curve predicted by SE; the finite $L$ transition follows very closely the asymptotic one. The two $R_{\rm{GAMP}}$ curves (dashed and solid) illustrate that, despite the mismatch in the rates between the Hadamard-based coding matrices and the Gaussian ones for low B, both rates coincide for large B. The region between the red and blue curves is the hard phase. To find $R^{\rm c}_{\rm{GAMP}}$, we follow the same procedure as for $R_{\rm{GAMP}}$ but using spatially coupled Hadamard-based operators and $L\!=\!2^{11}$. These are constructed as described in \cite{barbierSchulkeKrzakala,BarbierK15}, with the following coupling parameters, see middle inner figure for the block decomposition of a coupled coding operator: number of block-columns $L_c\!\in\!\{8, 16, 32\}$; number of block-rows $L_r\!=\! L_c\!+\!1$; backward and forward coupling windows $w_b\!\in\!\{2,3,5,7\}$, $w_f\!\in\!\{1,2\}$; coupling strength $\smash{\sqrt{J}}\!\in\![0.53,0.73]$ for the AWGNC, $0.3$ for the other channels (all the blocks other than the light blue coupling ones and the all-zeros blocks have unit strength); relative size of the ``seed'' block $\beta_{\rm seed}\!\in\![1.02, 1.25]$.}
	\label{fig:phaseDiag}
\end{figure}
\section{Numerical experiments}\label{sec:results}
In Fig.~\ref{fig:phaseDiag} we compare the optimal and GAMP performances in terms of attainable rate, denoted by $R_{\rm pot}$ and $R_{\rm GAMP}$ respectively. For all channels, there exists, as long as the noise is not ``too high'', a hard phase where GAMP is sub-optimal. Moreover, the use of Hadamard-based operators have a performance cost w.r.t Gaussian ones but which vanishes as $B$ increases; they both have the same algorithmic threshold for $B$ large enough (but still practical, $B \ge 64$ being enough).

Consider Gaussian matrices. An interesting feature is that in constrast with the AWGNC case \cite{BarbierK15}, $R_{\rm{GAMP}}$ for these binary input channels is not monotonously decreasing; it increases until some $B$ (that may be large) but, although it may be hard to observe numerically (except for the BEC), it then decreases to reach $\lim_{B\to\infty}R_{\rm GAMP}\!= \mathcal{F}(0|1)/(2\ln(2))\!\!<\!C$ \cite{barbierDiaMacris_thresholdSat}. However, a gap to capacity $C$ persists as long as spatial coupling is not employed.

Spatial coupling allows important improvements towards $R_{\rm pot}$ even in practical settings, confirming the universality of coupled SS codes under GAMP decoding as $\lim_{B\to\infty}R_{\rm pot}\!=\!C$ \cite{barbierDiaMacris_thresholdSat}. The mismatch between $R_{\rm pot}$ and $R^{\rm c}_{\rm GAMP}$ is due to finite size effects which are more evident in coupled codes (both chain lengths and coupling windows should go to infinity after $L$ for $R^{\rm c}_{\rm GAMP}$ to saturate $R_{\rm pot}$).
\section*{Acknowledgments}
We acknowledge Nicolas Macris, Florent Krzakala and Rüdiger Urbanke for helpful comments as well as Alper Kose and Berke Aral Sonmez for an early stage study of GAMP. This work was funded by the SNSF grant no. 200021-156672. 
\ifCLASSOPTIONcaptionsoff
\fi
\bibliographystyle{IEEEtran}
\bibliography{IEEEabrv,bibliography}
\end{document}